\documentclass[preprint2]{aastex}

\newcommand{\xmm}{{\it XMM-Newton}}

\newcommand{\psr}{\object{PSR J0737--3039}}

\shorttitle{Spectral Analysis of the Double Pulsar}
\shortauthors{E. Egron E., A. Pellizzoni, A. M. T. Pollock et al.}

\begin{document}


%%%%%\title{Spectral Analysis of The Two XMM-Newton Large\\
%%%%%       Programs of The Double Pulsar}

\title{LONG-TERM STUDY OF THE DOUBLE PULSAR J0737-3039 WITH XMM-NEWTON: SPECTRAL ANALYSIS}


\author{E. Egron\altaffilmark{1}, A. Pellizzoni\altaffilmark{1}, A. Pollock\altaffilmark{2,3}, M.~N. Iacolina\altaffilmark{1}, N. R. Ikhsanov\altaffilmark{4,5,6}, A. Possenti\altaffilmark{1}, M. Marongiu\altaffilmark{1}}
%\affil{INAF - Osservatorio Astronomico di Cagliari, Via della Scienza 5,
%    09047 Selargius (CA), Italy}
\email{egron@oa-cagliari.inaf.it}

\altaffiltext{1}{INAF - Osservatorio Astronomico di Cagliari, Via della Scienza 5,
      09047 Selargius (CA), Italy}
\altaffiltext{2}{European Space Agency, XMM-Newton Science Operations Centre, European Space Astronomy Centre, Apartado 78, 26891 Villanueva de la Canada, Madrid, Spain}
\altaffiltext{3}{Department of Physics and Astronomy, University of Sheffield, Hounsfield Road, Sheffield S3 7RH, England}
\altaffiltext{4}{Pulkovo Observatory, Pulkovskoe shosse 65-1, St. Petersburg, 196140 Russia}
\altaffiltext{5}{Saint-Petersburg State University, St. Petersburg, 198504 Russia}
\altaffiltext{6}{Special Astrophysical Observatory RAS, Nizhny Arkhyz, 369167 Russia}


\begin{abstract}
We present a long-term spectral monitoring of the unique Double Pulsar binary PSR J0737-3039 corresponding to two "Large Programs" performed by XMM-Newton in 2006 and 2011. 
%We confirmed that most of the observed emission ($>75\%$) is steadly produced by synchrotron and/or inverse Compton processes in the Pulsar A magnetosphere, although the corresponding X-ray pulsar efficiency would be unusually high ($1.3 \times 10^{31}$ erg s$^{-1} \sim\,0.2\%\,\dot{E}_{\mathrm{A}}$) assuming recent distance estimates ($\sim$ 1.1 kpc) for PSR J0737-3039. Thermal emission contributions are also required by the spectral fit and include a black-body component likely ascribed to Pulsar B (temperature $kT_{\mathrm{bb}} \sim 100-200$ eV and emission radius $< 200\,$m) and powered by Pulsar A's spin-down energy.
%No significant 
Spectral variability of pulsar emission in soft X-rays is not evident over 5 years, despite the significant relativistic spin precession in the considered time span ($\sim 25^{\circ}$).
We provide, for the first time, evidence of  hard X-ray emission from the system in the $5-8$ keV energy band. The standard spectral analysis was coupled to the energy dependent spatial analysis to confirm this excess, most likely ascribed to iron line emission.
%The standard spectral analysis was coupled to an independent source detection method based on likelihood spatial analysis.
%The Fe K$\alpha$ emission line at $6.4-6.97$ keV is widely considered as a distinct property of sources in which matter forms a disk around an accreting object. It has been detected in the X-ray spectra of many super-massive black holes and stellar-mass-black-hole, as well as in neutron-star X-ray binaries, and magnetic cataclysmic variables.
%Such a feature 
The Fe K$\alpha$ emission line at $6.4-6.97$ keV was previously unheard-of in non-accreting binary systems and could testify to the presence of a relic disk that survived the supernova explosions that terminated the lives of the Double Pulsar's stellar progenitors. The existence of a relic disk in this system reinforces speculation about the presence of similar structures around other peculiar classes of isolated neutron stars.
\end{abstract}
%(250 words).

%The Fe Kα emission line at 6.4-6.97 keV is widely considered as a distinct property of sources in which matter forms a disk around an accreting object. It has been detected in the X-ray spectra of many super-massive black holes and stellar-mass-black-hole, as well as in neutron-star X-ray binaries, and magnetic cataclysmic variables. 
%Long-term X-ray monitoring of the Double Pulsar binary by the XMM-Newton observatory shows evidence of emission in the 5-8 keV energy band, including a spectral excess ascribed to iron line emission. Such features were previously unheard-of in non-accreting binary systems and testify to the presence of a relic disk that survived the supernova explosions that terminated the lives of the Double Pulsar’s stellar progenitors. The existence of a relic disk in this system reinforces speculation about the presence of similar structures around other peculiar classes of isolated neutron stars.



\section{Introduction}

%Since its discovery in 2003,
% \citep{Lyne2004,Burgay2003} 
%the Double Pulsar PSR J0737-3039 %retains all the attention. 
Fourteen years after its discovery, the Double Pulsar PSR J0737-3039 is still
studied extensively.
%\textbf{remains one of the most surprising and fascinating system.}
%\textbf{has been well studied.}
This system is composed of two neutron stars: an old, fast, mildly recycled 22.7 ms pulsar (hereafter referred to as Pulsar A; \citealt{Burgay2003}) and a younger and much slower pulsar with a period of 2.77 s (hereafter Pulsar B; \citealt{Lyne2004}). The compact objects orbit each other in a very tight orbit in only 2.4 hr, with mean orbital velocities of about 1 million km/h \citep[see e.g.][for a review]{Kramer2008}. 

PSR J0737-3039 represents the most compact relativistic system and the only binary system in which both neutron stars have been detected as radio pulsars. The Keplerian and post-Keplerian parameters offer a unique test for theories of strong field gravity \citep{Kramer2008,Kramer2006,Lyne2004}, and constrain the masses of both neutron stars with high accuracy. The physical riches of the Double Pulsar have been highlighted from the radio observations \citep{Kramer2009,Kramer2008}: whereas the pulse profile of Pulsar A has been stable, the pulsed emission of Pulsar B showed strong orbital flux and profile variations, before disappearing entirely in 2008 \citep{Perera2010}. Moreover, as a direct consequence of the inclination of a system observed nearly edge-on (i$\sim89^{\circ}$), radio eclipses were detected when Pulsar A passed behind Pulsar B \citep{Breton2008,Kaspi2004,McLaughlin2004}. The study of the light curves testified to a new type of interaction between Pulsar A's relativistic particle wind and Pulsar B's magnetosphere in the previously unexplored close environment between the two neutron stars. This aroused the interest to carry out high-energy observations in the X-ray band to investigate further the intra-binary environment. 

Early observations performed by the Chandra and XMM-Newton satellites \citep{Pellizzoni2008,Possenti2008,Chatterjee2007,Campana2004,McLaughlin2004,Pellizzoni2004} pointed out the difficulty to constrain the origin of the multifold nature of the Double Pulsar's X-ray emission. The non-thermal pulsed emission from Pulsar A, although very soft (power-law slope $\Gamma \sim 3.3$), is clearly predominant in the X-ray flux \citep{Pellizzoni2008,Possenti2008,Chatterjee2007}. 

In the frame of a first XMM-Newton "Large program" of 235 ks exposure in 2006, pulsed X-ray emission from Pulsar B was also detected in part of the orbit. Due to its own low rotational energy loss, X-ray emission from Pulsar B can be only powered externally through the spin-down energy of Pulsar A \citep{Pellizzoni2008}. This emission, consistent with thermal radiation of temperature kT$\sim30$ eV and bolometric luminosity of $\sim 10^{32}$ erg s$^{-1}$, was ascribed to the heating of Pulsar B's surface by Pulsar A's wind. A hotter ($\sim 130$ eV) and fainter ($\sim 5 \times 10^{29}$ erg s$^{-1}$) thermal component, possibly originating from backfalling material heating the polar caps of either Pulsar A or Pulsar B, was also suggested from the spectral analysis by \citet{Pellizzoni2008}.

In 2011, a second deep XMM-Newton observation of 370 ks was carried out.
% yielding about 15,000 X-ray source counts in total, \textbf{considering EPIC-pn, MOS1 and MOS2 data from both large programs}. 
Comprehensive timing analysis over such a large time span has confirmed X-ray pulsed emission from Pulsar B even after its radio disappearance \citep{Iacolina2016}. The unusual phenomenology of Pulsar B's X-ray emission includes orbital pulsed flux and profile variations as well as a loss of pulsar phase coherence on time scales of years, suggesting orbital-dependent penetration of Pulsar A's wind plasma onto Pulsar B's closed field lines. Furthermore, timing analysis of the full XMM-Newton dataset provided first evidence of orbital flux variability ($7\pm1\%$), possibly involving a bow-shock between pulsar structures \citep{Iacolina2016}. An additional, possibly hard, spectral component associated with this intra-binary environment is then expected.

In this paper, we present
% the spectral analysis relative to the second large program of the Double Pulsar. Moreover, 
%we propose 
a comprehensive spectral analysis of the full XMM-Newton dataset from both large programs, which allows us for the first time to constrain high-energy components above 4 keV. Because of the weakness of the source and the consequent need to scrutinize the low-count statistics, an in-depth revision of the background subtraction procedure is provided. The standard spectral analysis was coupled with an independent source detection method based on likelihood spatial analysis.
The complementarity of both approaches allows us to speculate for the first time on the possible presence of an iron line in the Double Pulsar. 
%hard X-ray emission from the Double Pulsar. 

%In order to assess if systematic errors could affect the realm of such a hard component, we probed alternative methods described in literature, which provide comparable results. 

\section{Data reduction and background mitigation}

%A first large program on PSR J0737–3039 was carried out by XMM-Newton in October 2006 for a total exposure time of ∼235 ks (XMM archive observation IDs 0405100101/0201/0401). Five years later, a second large program made possible the observation of the Double Pulsar during three XMM-Newton orbits, which corresponds to the coverage of 41 binary system orbits, roughly $50\%$ more than during the first large program (observation IDs 0670810101/0201/0301) for a total exposure time of 370 ks. 

\begin{deluxetable}{clcccccc}
\tabletypesize{\small}
%\rotate
\tablecaption{Total exposure time and observing time of both large programs\label{TabExpTime}}
\tablewidth{0pt}
%\tablenum{1}
\tablehead{
\colhead{Obs date} & \colhead{Instrument} & \multicolumn{3}{c}{Total Exp Time} & \multicolumn{3}{c}{Observing Time\tablenotemark{a}} \\
& & & (ks) & & & (ks)  \\
& & \it{orb1\tablenotemark{b}} & \it{orb2}  & \it{orb3}  & \it{orb1}  & \it{orb2}  & \it{orb3} 
}
\startdata
 %  2006 & pn        & 119.8 + 114.5 & 82.3 + 76.1 \\
  2006 & pn        & 120 & 115 &  &  82 & 76 \\
 %           & mos\,1 & 119.9 + 114.3 & 111.8 + 103.0 \\ 
           & MOS\,1 & 120 & 114 & &  112 & 103 \\ 
 %           & mos\,2 & 119.9 + 114.3 & 112.2 + 103.9 \\
           & MOS\,2 & 120 & 114 & & 112 & 104 & \\
%   2011 & pn & 129.3 + 106.1 + 128.3 & 87.8 + 67.7 + 84.1\\
  2011 & pn & 129 & 106 & 128 & 88 & 68 & 84\\
%            & mos\,1 &  107.7 + 103.4 + 128.4 & 104.4 + 83.9 + 111.8 \\
           & MOS\,1 &  108 & 103 & 128 & 104 & 84 & 112 \\
%            & mos\,2 &  129.4 + 103.4 + 128.4 & 122.2 + 84.2 + 112.2 \\
           & MOS\,2 &  129 & 103 & 128 & 122 & 84 & 112 \\
\enddata
%% Text for table notes should follow after the \enddata but before
%% the \end{deluxetable}. Make sure there is at least one \tablenotemark
%% in the table for each \tablenotetext.
%\tablecomments{See text in Sec.\,2 for details on data reduction and background screening. 
%The reported values have the following selection: pattern $\le 4$ for the EPIC-pn and pattern $\le 12$ for the EPIC-mos.
%}
\tablenotetext{a}{After dead-time correction and screening for soft proton flares.
%, with pattern $\le 4$ for the EPIC-pn and pattern $\le 12$ for the EPIC-mos.
}
\tablenotetext{b}{``\textit{orb*}'' refers to the \xmm\ orbits during the observations of \psr: orbits \textit{1260} and \textit{1261} in 2006, and orbits \textit{2174, 2175} and \textit{2176} in 2011.}
\end{deluxetable}

Both XMM-Newton Large Programs were carried out with similar instrumental configurations suitable for simultaneous spectral and high-resolution timing analysis of the Double Pulsar. For these purposes, the EPIC-pn \citep{Struder2001} was used in “Small Window” mode whose time resolution of 5.67 ms compares with Pulsar A's period of 22.7 ms. The EPIC-MOS cameras \citep{Turner2001} were also operated in “Small Window” mode with a time resolution of 0.3 s only suitable for Pulsar B timing analysis. Medium and thin optical filters were applied to the EPIC instruments in 2006 and 2011, respectively. The thin filter nearly doubles the instrument effective area in the soft band ($0.3-0.5$ keV) at the expense of higher background contamination: early X-ray observations of the Double Pulsar showed a very soft source spectrum motivating the use of the thin filter for subsequent observations in order to improve the overall counting statistics.

The XMM-Newton data were processed using the Science Analysis Software (SAS) version 12. The calibrated and concatenated EPIC event lists were obtained by running the meta-tasks epproc and emproc, standard pipeline tasks for EPIC-MOS and EPIC-pn observations respectively. We performed a barycentric correction on the event files thanks to the task barycen using the JPL DE405 ephemeris to convert the time of arrival of photons from the satellite position to the solar-system barycenter.

Because of the faintness of the Double Pulsar in X-rays, careful attention was provided in the assessment and mitigation of background contamination and in the selection of source photons.
Strong flares most likely associated with cosmic soft proton events, significantly affected different observations. In order to discard these episodes, we produced the light curves between 10 and 12 keV to determine the flaring times.
We followed a background rejection procedure according to the general prescriptions by \citet{DeLuca2004} and further investigating and testing differing filtering options in analogy with the espfilt SAS task, and reviewing the procedure adopted by \citet{Iacolina2016}  and \citet{Pellizzoni2008} for the same data. Two different photon pattern selections were considered for the distribution of incident photons over the pixels of the EPIC-pn (carrying out the analysis for both options): single and double events (pattern $\leq 4$) or single events only (pattern = 0), slightly reducing the effective area. As for the EPIC-MOS, single, double, triple and quadruple events were selected (pattern $\leq12$).

\begin{deluxetable}{ccccccccccc}
\tabletypesize{\scriptsize}
%\rotate
\tablecaption{Comparison of three models on the second large program data\label{Comp3-largeprog2008}}
\tablewidth{0pt}
%\tablenum{2}
\tablehead{
\colhead{Detector} & \colhead{Model\tablenotemark{a}} & \colhead{$N_{H}$} & 
\colhead{$\Gamma$} & \colhead{Norm$_{po}$} & \colhead{$kT_{bb1}$} & \colhead{Norm$_{bb1}$} & \colhead{$kT_{bb2}$} & \colhead{Norm$_{bb2}$} & \colhead{$\chi^{2}/$dof} & \colhead{$\chi^{2}_{red}$}\\
 & & ($10^{20}$ cm$^{-2}$) & &  & (eV) & & (eV) & & 
}
\startdata
pn & 1 & $0.8^{+1.2}_{-0.7}$ & $2.9 \pm 0.3$ & $5.3 \pm 1.2$ & $150^{+20}_{-10}$ & $2.0 \pm 0.6$ & & & 239/206 & 1.16 \\

     & 2 & -& & & $100 \pm 8$ & $3.8 \pm 0.3$ & $250 \pm 30$ & $2.0^{+0.4}_{-0.3}$ & 254/206 & 1.23 \\

     & 3 & - & $2.6 \pm 0.4$ & $3.4^{+1.2}_{-1.4}$ & $110 \pm 30 $ & $2.1^{+0.8}_{-1.9}$ &  $210^{+160}_{-100}$ & $1.1^{+1.1}_{-0.9}$ & 237/204 & 1.16 \\
pn+MOS & 1 & $1.5 \pm 0.9$ & $3.1 \pm 0.2$ & $ 6.0\pm 0.9$ & $160^{+20}_{-10}$ & $1.8 \pm 0.4$ & & & 355/318 & 1.12 \\

    	     & 2 & - & & & $105 \pm 6$ & $3.7 \pm 0.2$ & $260 \pm 20$ & $1.9 \pm 0.2$  & 367/318 & 1.15\\

    	     & 3 & $\leq$ 1.0 & $2.7^{+0.4}_{-0.3}$ & $2.6 \pm 0.1$ & $110 \pm 20$ & $2.4^{+0.5}_{-0.8}$ & $230^{+50}_{-40}$ & $1.3 \pm 0.5$ &  346/316 & 1.10 \\

%tbabs & $N_{H}$ & $10^{20}$ cm$^{-2}$ & $1.7\pm 0.9 &   \\
%bbody & $kT$ & eV & $147^{+20}_{-13}$ &  \\
%bbody & $Norm$ & $\times 10^{-7}$ & 1.6 \pm 0.5 \\
%po & $\Gamma$ & & $3.0 \pm 0.2}$ &   \\
%po & $Norm $ & & $6.1 \pm 0.9$ &  \\
 %  $\chi^{2}/$dof & & &347/333   \\
  %$\chi^{2}_{red}$ & & & 1.04  \\
\enddata
%% Text for table notes should follow after the \enddata but before
%% the \end{deluxetable}. Make sure there is at least one \tablenotemark
%% in the table for each \tablenotetext.
%\tablecomments{The ``-'' in the $N_{H}$ column indicates that the value is found to be very low and not constrained.}
\tablenotetext{a}{Models: 1) \textsc{tbabs}*(\textsc{po}+\textsc{bb}), 2) \textsc{tbabs}*(\textsc{bb}+\textsc{bb}), 3) \textsc{tbabs}*(\textsc{po}+\textsc{bb}+\textsc{bb}). }
\tablenotetext{b}{The power-law and black-body normalizations are in units of $10^{-6}$ photons cm$^{-2}$ s$^{-1}$ keV$^{-1}$ and $10^{-7}$, respectively.}
\tablenotetext{c}{The ``-'' in the $N_{H}$ column indicates that the value is found to be very low and not constrained.}
\end{deluxetable}

  \begin{figure}
  \begin{center}
  \begin{tabular}{c}
%%\begin{figure*}
%        \includegraphics[height=6.2cm]{Fig-small-window_PN.pdf}
        \includegraphics[height=6.2cm]{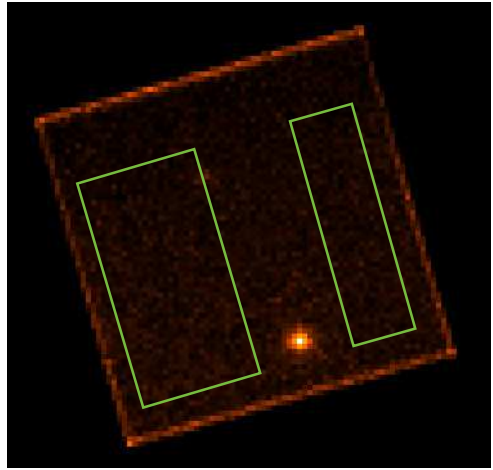}
       \end{tabular}
   \end{center}
   \caption{\small{Small-window EPIC-pn image corresponding to the 2006 data of the Double Pulsar (first XMM-Newton orbit). The bright spot indicates the target, while the green rectangles shows the areas selected to extract the background spectrum.}}
    \label{fig:small-wind-PN}
%%\end{figure*}
 \end{figure}

In parallel, we cross-checked our analysis by adopting another method, which consists in analyzing the average quiescent counts rates free from flaring particle background. We extracted a 10\,s binned light curve in the $0.15-10$ keV energy range from the whole field of view of each camera, omitting the source region, and selecting events below a threshold of $5\sigma$ of the quiescent rate for the EPIC-pn for pattern = 0 selection, and $4\sigma$ for pattern $\leq 4$. As for the EPIC-MOS, the threshold was defined at $5\sigma$ from the quiescent count rate. The resulting effective observing time after this operation is reported in Table~\ref{TabExpTime} together with the original total exposure time for each EPIC instrument (no significant dead-time differences are related to the two photon pattern options considered for the EPIC-pn). Alternative photon pattern selections and threshold levels provide consistent results, although the choice of  pattern $\leq 4$ for the EPIC-pn is best suited for hard X-ray spectral analysis providing a higher effective area at the expenses of a background increase only in the softer energy channels ($< 0.4$ keV). 

Once the EPIC event lists were cleaned from strong background contamination, we extracted the source and background spectra. The associated source's radius of extraction was set at $18\arcsec$ for the EPIC-pn, and $15\arcsec$ for the EPIC-MOS providing maximum source detection significance (pulsed emission from Pulsar A) and carefully checking that residual flare contamination was not present within the selected region. The background spectra were extracted from different rectangular regions for consistency checks (see Fig.~\ref{fig:small-wind-PN}). The task backscale was used to calculate the area of source and background regions. Standard energy response matrices (RMFs; rmfgen) and effective area files (ARFs; arfgen) were created for each spectrum via the standard SAS tools.
The background filtering and source extraction procedures provided an overall harvest of $\sim 15,000$ source X-ray photons considering the EPIC-pn, MOS1 and MOS2 data from both large programs.

\section{Spectral analysis}

Up to five significant spectral components could be in principle expected from the Double Pulsar in X-rays accordingly to theoretical models: surface thermal emission and non-thermal magnetospheric emission from both neutron stars, and orbital-phase dependent bow-shock emission due to the interaction between Pulsar A's wind and Pulsar B's magnetosphere.  
This latter non-thermal component, if present, was proved to be much weaker than expected: no clear evidence of shock emission was reported in the spectral analysis of 2006 data. %\citep{Pellizzoni2008} .
%and the 2011 data. 
%The spectrum appeared clearly dominated by pulsed non-thermal emission from Pulsar A 
While the spectrum related to that data set was correctly fitted by a two-component model (a power law plus a black body or two black bodies), the phase resolved analysis indicated that the spectrum appeared clearly dominated by pulsed non-thermal emission from Pulsar A and the off-pulse required two-components thermal emission models \citep{Pellizzoni2008}.
Furthermore, only very weak orbital flux modulation ($7\%$) was detected as a possible signature of a bow-shock in the overall timing analysis of both large programs \citep{Iacolina2016}. 
Thus, a two (pl+bb) or three-component (pl+bb+bb) pulsar emission model  could satisfactorily account for most of the X-ray flux from the Double Pulsar. 
%\textbf{Thus three-component model (pl+bb+bb) could satisfactorily account for the X-ray flux from the Double Pulsar. }
%We verified that the spectral analysis of the observations performed in 2011, as well as combining 2006+2011 data, provides results comparable with those related to 2006 data alone (i.e. no significant spectral variability is evident over years). \textbf{The parameters associated to the different models are reported in Table~\ref{Comp3-largeprog2008}...

We used the XSPEC version 12.8 \citep{Arnaud1996}. All uncertainties are given at the $90\%$ confidence level ($\Delta\chi^2 =2.706$). The EPIC-pn, MOS1 and MOS2 data were rebinned to have at least 25 counts per energy channel.
The bulk of the X-ray emission from PSR J0737-3039 is at low energy, below $\sim$1 keV.
% from 0.15 to 3 keV.
However, in order to better constrain the models, the spectral analysis was performed in the $0.15-11$ keV energy range.
%For the sake of completeness, we tested both two-component and three-component fits as suggested by the phase-resolved studies on the 2006 data \citep{Pellizzoni2008}. 
%The models were first applied to the 2011 data set, then to the overall data set.
%although the former were disfavoured by previously reported phase-resolved studies (Pellizzoni et al. 2008). 
%The probed spectral scenarios were 
The probed models were all modified at low energies to account for interstellar photoelectric absorption through XSPEC Tbabs model \citep{Wilms2000}. Tbabs calculates the cross section for X-ray absorption by the ISM as the sum of the cross sections for X-ray absorption due to the ISM in its gas, grain and molecular phases. 

We first verified that the spectral analysis of the observations performed in 2011 
%as well as combining 2006+2011 data, 
provides results comparable with those related to 2006 data alone (i.e. no significant spectral variability is evident over years). For the sake of completeness, we tested both two-component and three-component fits as suggested by the phase-resolved studies on the 2006 data \citep{Pellizzoni2008}. 
The models were applied on the EPIC-pn data, then EPIC-pn plus MOS1 and MOS2 data.
The parameters associated to the different models are reported in Table~\ref{Comp3-largeprog2008}.
%
%We applied the different models on the 2011 data first. 
%We tested a two-component model made up of a power law and a black body.
%The associated parameters are $N_{H} \sim (1.5\,\pm\,0.9) \times 10^{20}$ cm$^{-2}$, the photon index of the power law $\Gamma = 3.1\,\pm\,0.2$, and the temperature of the black
%body $kT_{bb}=160^{+20}_{-10}$ eV. The corresponding $\chi^{2}$ is 355/318 ($\chi^{2}_{red}\,\sim\, 1.12$).
%These values are perfectly in agreement with the first XMM-Newton large program's data of PSR J0737-3039 performed in 2006 \citep{Pellizzoni2008}. The observed flux in the $0.2-3$ keV energy range is $\sim 4.3 \times 10^{-14}$ erg/cm$^{2}$/s. By comparison, the observed flux was estimated at $\sim 4.0 \times 10^{-14}$ erg/cm$^{2}$/s during the observations in 2006.
% \citep{Pellizzoni2008}.
The results obtained separately for the two large programs are perfectly in agreement (see \citealt{Pellizzoni2008}). 
%We note that the hydrogen column density is more difficily constrained during the second large program. 
The observed flux related to the 2011 data in the $0.2-3$ keV energy range is $\sim 4.3 \times 10^{-14}$ erg/cm$^{2}$/s (using the power law plus black body model). By comparison, the observed flux was estimated at $\sim 4.0 \times 10^{-14}$ erg/cm$^{2}$/s during the observations in 2006.
We therefore analyzed both data sets together.
%We first tested a power law plus a black-body model on the EPIC-pn and EPIC-MOS data separately, and then coupled together. The associated $\chi^2$ is 533 for 520 dof. The observed flux in the $0.2-3$ keV energy range is $4.3\times10^{-14}$ erg cm$^{-2}$s$^{-1}$. The addition of a second black body to this model gives a $\chi^2=524$ for 518 dof. We then deleted the power-law in order to test a two-blackbody model. The reduced $\chi^2$ is still good, but the hydrogen column density is not constrained. By forcing this parameter to have the same value as in the previous models, the $\chi^2$ increases quite significantly ($\chi^2=615$). The null hypothesis probability associated with the different models is $17\%$ for tbabs*(bb+bb), $35\%$ for tbabs*(po+bb), and $41\%$ for tbabs*(po+bb+bb). 
A summary of models parameters is reported in Table~\ref{Comp3}.

%%%%%%%%%%%%%%%%%%%%%%%%%%

\begin{deluxetable}{ccccccccccc}
\tabletypesize{\scriptsize}
%\rotate
\tablecaption{Comparison of three models using the data of both large programs\label{Comp3}}
\tablewidth{0pt}
%\tablenum{2}
\tablehead{
\colhead{Detector} & \colhead{Model\tablenotemark{a}} & \colhead{$N_{H}$} & 
\colhead{$\Gamma$} & \colhead{Norm$_{po}$} & \colhead{$kT_{bb1}$} & \colhead{Norm$_{bb1}$} & \colhead{$kT_{bb2}$} & \colhead{Norm$_{bb2}$} & \colhead{$\chi^{2}/$dof} & \colhead{$\chi^{2}_{red}$}\\
 & & ($10^{20}$ cm$^{-2}$) & &  & (eV) & & (eV) & & 
}
\startdata
pn & 1 & $1.7\pm 0.9$ & $3.0 \pm 0.2$ & $6.1 \pm 0.9$ & $147^{+20}_{-13}$ & $1.6 \pm 0.5$ & & & 347/333 & 1.04 \\

     & 2 & -& & & $105 \pm 5$ & $3.9 \pm 0.2$ & $260^{+25}_{-20}$ & $1.8^{+0.3}_{-0.2}$  & 360/333 & 1.08 \\

     & 3 & $\leq$ 1.7 & $2.5^{+0.7}_{-0.6}$ & $2.7^{+2.8}_{-1.3}$ & $110 \pm 20$ & $2.7^{+0.7}_{-1.5}$ &  $230^{+90} _{-55}$ & $1.2^{+0.7} _{-0.6}$ & 343/331 & 1.04 \\

pn+MOS & 1 & $2.2 \pm 0.7$ & $3.2 \pm 0.2$ & $6.1 \pm 0.7$ & $160 \pm 15$ & $1.6^{+0.4}_{-0.3}$ & & & 533/520 & 1.02 \\

    & 2 & - & & & $105 \pm 5$ & $3.9 \pm 0.2$ & $270 \pm 20$ & $1.8 \pm 0.2$  & 551/520 & 1.06\\

     & 3 & $\leq$ 1.6 & $2.5^{+0.7}_{-0.5}$ & $2.3^{+3.2}_{-0.9}$ & $110\pm 10$ & $2.8^{+0.5}_{-1}$ & $230^{+40}_{-30}$ & $1.4^{+0.5}_{-0.7}$ & 524/518 & 1.01 \\

%tbabs & $N_{H}$ & $10^{20}$ cm$^{-2}$ & $1.7\pm 0.9 &   \\
%bbody & $kT$ & eV & $147^{+20}_{-13}$ &  \\
%bbody & $Norm$ & $\times 10^{-7}$ & 1.6 \pm 0.5 \\
%po & $\Gamma$ & & $3.0 \pm 0.2}$ &   \\
%po & $Norm $ & & $6.1 \pm 0.9$ &  \\
 %  $\chi^{2}/$dof & & &347/333   \\
  %$\chi^{2}_{red}$ & & & 1.04  \\
\enddata
%% Text for table notes should follow after the \enddata but before
%% the \end{deluxetable}. Make sure there is at least one \tablenotemark
%% in the table for each \tablenotetext.
%\tablecomments{The ``-'' in the $N_{H}$ column indicates that the value is found to be very low and not constrained.}
\tablenotetext{a}{Models: 1) \textsc{tbabs}*(\textsc{po}+\textsc{bb}), 2) \textsc{tbabs}*(\textsc{bb}+\textsc{bb}), 3) \textsc{tbabs}*(\textsc{po}+\textsc{bb}+\textsc{bb}). }
\tablenotetext{b}{The power-law and black-body normalizations are in units of $10^{-6}$ photons cm$^{-2}$ s$^{-1}$ keV$^{-1}$ and $10^{-7}$, respectively.}
\tablenotetext{c}{The ``-'' in the $N_{H}$ column indicates that the value is found to be very low and not constrained.}
\end{deluxetable}

Since the hypothetical presence of significant bow shock emission should be associated with a hard photon index \citep{Sturner1997}, we tried to fix the photon index of the power law to a lower value, such as $\Gamma \sim 2$. The resulting fit leads to a $\chi^2_{red} \sim 1.34$. 
%and no constraints on the NH any- more. 
These results are also consistent with \citet{Pellizzoni2008}.

The reported weak orbital flux variability of $7\%$ from timing analysis 
\citep{Iacolina2016} could be in principle constrained by orbital phase resolved analysis, 
in spite of the relatively low count statististics involved. In fact, splitting the orbital
phase interval in different sections, we did not find any significant variation of the above spectral components.
In particular, we cannot claim any significant spectral changes in 
correspondence to peculiar orbital phases (around neutron star 
conjunctions and quadratures, and periastron/apoastron passages) where putative X-ray 
emission could in principle be enhanced due, for example, to an hypothetical anisotropic bow-shock emission \citep{Lyutikov2004,Granot2004}.

\begin{figure}
   \begin{center}
        \begin{tabular}{c}
       \includegraphics[height=6.3cm]{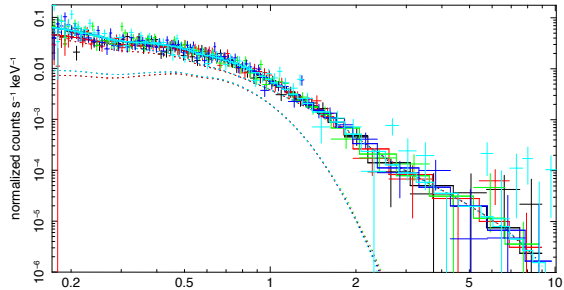}
       \end{tabular}
   \end{center}
   \caption{\small{Two-component model (power law plus black body) applied on the EPIC-pn data obtained in 2006 and 2011 (pattern=0). The corresponding values of the parameters are reported in Table\,~\ref{Comp3}. The different colors refers to the different \xmm\ orbits (black and red: 2006, green, blue and cyan: 2011)}.
%Comparison of the three models applied on the EPIC-pn data obtained in 2006 and 2011 (pattern=0). The corresponding values of the parameters are reported in Table\,~\ref{Comp3}. The different colors refers to the different \xmm\ orbits (black and red: 2006, green, blue and cyan: 2011)}.
}
    \label{fig:Comp3Model}
 \end{figure}

Though perfectly adequate for the soft X-ray band, none of these two or three-component models properly fits the data above 4 keV. %(see Fig.~\ref{fig:Comp3Model}).
By applying a two or three-component model from Table~\ref{Comp3}, an excess is clearly observed at $6-7$ keV (see Fig.~\ref{fig:Comp3Model}), while error bars become very important above 8 keV where the background dominates.
%%%Indeed, a spatial model (see the following subsection) applied on $8-10$ keV confirms that photons are below any sensible detection threshold in this energy range. 
%We added a narrow Gaussian line to fit the spectral feature. Considering the standard patterns $\leq\,4$, the line is found at $6.2\,\pm\,0.5$ keV. The associated line width is constrained to $\sigma <\,0.8$ keV and the flux is $(3 \pm 2)\times 10^{-7}$ photons cm$^{-2}$s$^{-1}$keV$^{-1}$.

\begin{figure}
   \begin{center}
        \begin{tabular}{c}
        \includegraphics[height=6.2cm]{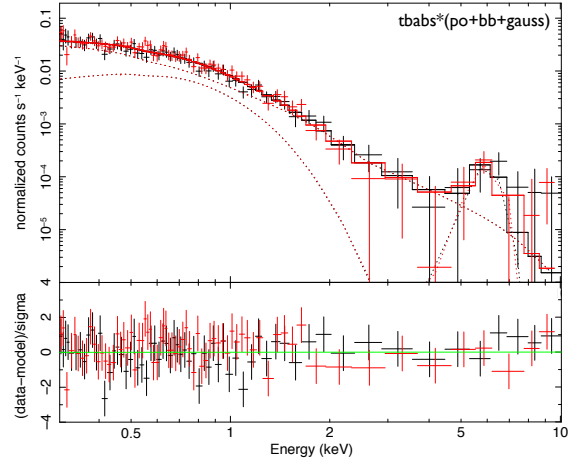}
       \end{tabular}
   \end{center}
   \caption{\small{Addition of a Gaussian line at $\sim 6.5$ keV to fit the high-energy photons in the EPIC-pn data obtained in 2006. The background dominates the spectra at E $\geq 8$ keV.}}
    \label{fig:PnFe2006Gauss}
 \end{figure}

The EPIC-pn data obtained during the two Large Programs were studied separately in order to assess the spectral parameters of this newly-recognized hard component, including possible variability. The spectrum obtained in 2006 indicates an excess at $6-8$ keV in the two XMM-Newton orbits.
%, as shown in Fig.~\ref{fig:PN-Fe-2006}. 
A narrow Gaussian line fits adequately this excess (see Fig.~\ref{fig:PnFe2006Gauss}). Its centroid energy is found at $6.2\,\pm\,0.5$ keV with the associated line width constrained to $\sigma<\,0.8$ keV. The line flux $(3 \pm 2) \times 10^{-7}$ photons cm$^{-2}$ s$^{-1}$ keV$^{-1}$ represents $7\%$ of the total absorption-corrected luminosity L$_{0.3-11\mathrm{keV}}$ = $6.1\,\pm\,0.3 \times 10^{30}$ erg/s (assuming a distance of 1.1 kpc; \citealt{Verbiest2012}). 
%consistent with the flux independently detected through likelihood analysis. 
No other simple model, such as a power law or a black body, is able to fit the high-energy excess. 
%Emission at these energies indicates the likely presence of iron line emission. The combined probability of a fake detection of hard X-ray photons through both spectral analysis and spatial analysis is negligible ($\sim\,3 \times 10^{-6}$).
%%%%%%%%%%%%%%%%%%%%%
Since only a few source photons are detected at high energy ($\sim 1\%$ in the $4-8$ keV range), it is not possible to assess the improvement of the fit for the whole energy range including the Gaussian line. A spectral fit comparison with or without the line was performed on the restricted energy range $4-8$ keV. A power law plus a blackbody component were applied to the data, with the parameters fixed to the best values found in the whole energy range. The associated $\chi^2$ is 11.4 for 7 dof ($\chi^{2}_{red} \sim 1.6$), indicating that an additional spectral contribution at high-energy is required. Indeed, the addition of a Gaussian line at $\sim 6.2$ keV significantly improves the fit, reaching $\chi^2 = 3.4$ for 4 dof ($\chi^{2}_{red} \sim 0.85$). Comparable results are obtained assuming other models in Table~\ref{Comp3}.

The use of the C-statistic \citep{Cash1979} instead of the $\chi^2$ is in principle more appropriate for the fitting procedure with only a few photons per energy bin in the spectra, although its exploitation in the frame of XSPEC is not straightforward. Since the background dominates at energy $> 8$ keV, we restricted the analysis in the energy range $0.15-8$ keV when considering pattern = 0, and $0.3-8$ keV in the case of pattern = $0-4$. The previously discussed model was applied to the data, -i.e. a power law plus a blackbody component. 
Adopting the C-statistic and pattern selection $\leq 4$, the Gaussian line centroid is found at $6.1\,\pm\,0.3$ keV, with line width $\sigma<\,0.8$ keV and normalization $\sim 3\times10^{-7}$ photons cm$^{-2}$ s$^{-1}$ keV$^{-1}$, in agreement with results obtained through the use of $\chi^2$ statistic. As for the case of the $\chi^2$ statistic, in order to quantify the improvement of the fit provided by the introduction of a Gaussian component, we compared the values of the C-statistic using continuum models in the restricted $4-8$ keV band, with or without the inclusion of the line. By applying only the continuum to the data, a C-statistic value of 195 for 182 dof indicates a bad fit. The addition of the Gaussian at 6.2 keV definitely improves the C-statistic providing 179.4 for 179 dof. Comparable results are obtained adopting the strictest criterion photon pattern=0.
Since our procedure can be assimilated to an F-test which is not strictly adequate to assess the statistical significance of the line \citep{Protassov2002},
%ApJ 571, 545), 
we also provided spectral simulations of fake data based on power law plus black-body models described above. These simulations show that (1) background counts dominate at $E> 8$ keV, (2) continuum 
emission counts drops above 5 keV. Chance probability of a fluctuation of source continuum emission mimicing a line at $6-7$ keV appears negligible. Even after 400 trials, fake continuum 
data never required the inclusion of a line feature in the fit, confirming line 
detection above 3-sigma level.

\begin{figure}
   \begin{center}
        \begin{tabular}{c}
    \includegraphics[height=6.2cm]{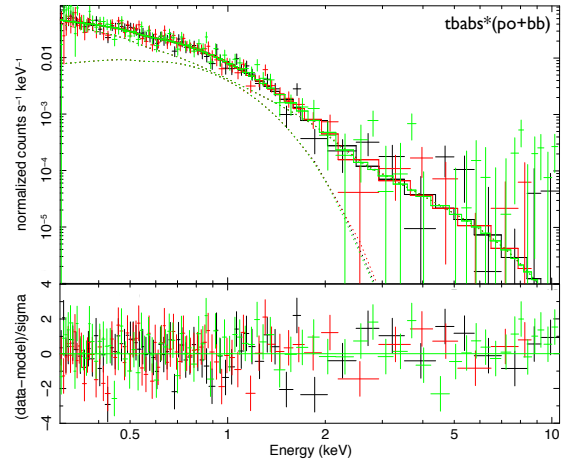}
       \end{tabular}
   \end{center}
   \caption{EPIC-pn data obtained in 2011 with the selection of pattern $\leq 4$. Above 2 keV, photons are not well fitted by the model.
}
    \label{fig:Pn2011tail}
 \end{figure}

In the case of the second long observation performed in 2011, the high-energy component appears to be different from that detected in 2006, with weaker evidence of a high-energy spectral line
%Instead of a bump, it appears as a tail beginning at 2.5 keV 
(see Fig.~\ref{fig:Pn2011tail}). 
However, it is worth noting that the quality of the data is not as good as during the observation performed in 2006, with a stronger background emission partly due to the use of the EPIC “thin” filter. 
We simulated fake data including a continuum (power-law plus black-body) model and a line feature with the same parameters derived from 2006 data, assuming the background model obtained from 2011 data. An emission line with the same energy and flux as in the 2006 observation would appear at the limit of detectability in 2011 data.
%Careful simulations proved that an emission line with the same energy and flux as in the previous observation would be at the limit of detectability. 
%However, it is worth noting that the data corresponding to the second XMM orbit are more different from the other ones. It could be related to stronger high particle background episodes identified in the $10-12$ keV light-curve. However, even using a lower threshold at $3\sigma$ from the quiescent rate to screen from these soft proton flares, this observation remains different.
%We then considered only the first and the third XMM orbits in 2011 plus the observations per- formed in 2006. The bump at $6-7$ keV is visible when pattern = 0 or pattern = $0-4$ are selected, but obviously, the addition of a Gaussian line does not improve the fit. The centroid energy is found at $\sim5.5$ keV and the $\sigma$ value is very high. The bump is still visible even with the addition of the line, confirming that the data obtained in 2006 and 2011 present a different high-energy component. We discuss the issues further in \S5.2.

Unfortunately, orbital phase-resolved analysis did not provide any significant constraints at E  $> 4-5$ keV because of the low counts statistics, as well as pulsar phase-resolved analysis was not suitable in hard X-rays.

%\begin{figure}
%   \begin{center}
%        \begin{tabular}{c}
%        \includegraphics[height=6.2cm]{Fig2_Fe.pdf}
%       \end{tabular}
%   \end{center}
%   \caption{\small{EPIC-pn data obtained in 2006  in the case of a selection of pattern $\leq 4$. An excess is clearly visible at E $>$ 4 keV. The background dominates the spectra at E $\geq 8$ keV.}}
%    \label{fig:PN-Fe-2006}
% \end{figure}

\section{High-energy spatial analysis}

In order to exclude that the reported hard X-ray features are  associated with background emission, the standard spectral analysis was coupled to an \textit{independent} source detection method based on likelihood spatial analysis. The most sensitive procedure to estimate the strength of sources in X-ray data in common with other photon-counting data entails constructing composite models of the spatial distribution of events combining source and background components \citep{Pollock1987}. The background near the Double Pulsar is made up of cosmic and instrumental contributions and is assumed to be flat. 

A point spread function spatial model was applied to EPIC images between 4 and 8 keV, where the discrepancy between the low-energy spectral models and the hard X-ray component is the most relevant. The Table~\ref{spatial-table} reports the results separately and combined for each of the 5 long observations comprising the two Large Programs in 2006 and 2011. The exposure time reported, T(s), includes the dead-time correction of the Small Window mode. The likelihood detection statistic, lnL, shows that the Double Pulsar is a weak source at high energies that is difficult to detect in a whole spacecraft revolution of nearly two days, but which is consistent in strength with the count rate accumulated in half a million seconds of elapsed time.

The combined log-likelihood detection statistic of 15.1 is a secure $\sim 4\sigma$ detection that amounts to $160 \pm 40$ photons above 4 keV from the X-ray source. Unlike spectral analysis, unpredictable background contamination cannot significantly affect likelihood analysis results, since background counts are not distributed in the image accordingly to the source point spread function. 
Moreover, the spatial analysis testifies that there is no detection of source photons above 8 keV.
Therefore our analysis proves that hard X-ray photons originate from the Double Pulsar in the $4-8$ keV energy range.

%Furthermore, the standard spectral analysis was coupled to an independent source detection method based on likelihood spatial analysis. A point spread function spatial model was applied to EPIC images between 4 and 8 keV, where the discrepancy between the low-energy spectral models and the hard X-ray component is the most relevant. The combined log-likelihood detection statistic of 15.1 is a secure 4σ detection that amounts to 160±40 photons above 4 keV from the X-ray source. Unlike spectral analysis, unpredictable background contamination cannot significantly affect likelihood analysis results, since background counts are not distributed in the image accordingly to the source point spread function. The complementarity of the independent spatial and spectral approaches allows us to claim for the first time hard X-ray emission from the Double Pulsar. 

%Sources are modelled with the Point-Spread-Function included among the calibration files supplied by the XMM-Newton Observatory. In the search for high-energy emission from the Double Pulsar, the analysis was performed between 4 and 8 keV with the EPIC-pn, the most sensitive of the XMM-Newton instruments at high energies. 

\begin{deluxetable}{cccccccc}
%\tabletypesize{\scriptsize}
\tabletypesize{\small}
%\rotate
\tablecaption{Likelihood detection statistic associated with the five XMM-Newton orbits \label{spatial-table}}
\tablewidth{0pt}
%\tablenum{2}
\tablehead{
\colhead{Obs. ID} & \colhead{MJD start} & \colhead{MJD stop} & 
%\colhead{x} & \colhead{y} & 
\colhead{T(s)} & \colhead{Count Rate (/s)} & \colhead{lnL} %\colhead{Norm$_{bb2}$} & \colhead{$\chi^{2}/$dof} & \colhead{$\chi^{2}_{red}$}\\
% & & ($10^{20}$ cm$^{-2}$) & &  & (eV) & & (eV) & & 
}
\startdata
0405100201 & 54034.037 & 54035.423 &  82710 & $0.0003\,\pm\,0.0001$ & 3.0 \\
0405100101 & 54036.037 & 54037.362 &  79067 & $0.0007\,\pm\,0.0003$ & 3.2 \\
0670810201 & 55856.863 & 55858.360 &  89353 &  $0.0003\,\pm\,0.0001$ & 3.1 \\
0670810301 & 55858.828 & 55860.056 & 73249  & $0.0010\,\pm\,0.0003$  & 5.1 \\
0670810101 & 55860.854 & 55862.339 & 88675 & $0.0005\,\pm\,0.0002$  &  3.8  \\
%
%\hline
\textbf{Combined}      &           &            & \textbf{413055}  & \textbf{0.00038}$\,\pm\,$\textbf{0.0002} & \textbf{15.1}  \\
\enddata
%% Text for table notes should follow after the \enddata but before
%% the \end{deluxetable}. Make sure there is at least one \tablenotemark
%% in the table for each \tablenotetext.
\tablecomments{XMM Epic-PN in the $4-8$ keV energy range. "Combined" corresponds to the combination of the data.}
%\tablenotetext{a}{Models: 1) \textsc{tbabs}*(\textsc{po}+\textsc{bb}), 2) \textsc{tbabs}*(\textsc{bb}+\textsc{bb}), 3) \textsc{tbabs}*(\textsc{po}+\textsc{bb}+\textsc{bb}). }
%\tablenotetext{b}{The power-law and black-body normalizations are in units of $10^{-6}$ photons cm$^{-2}$ s$^{-1}$ keV$^{-1}$ and $10^{-7}$, respectively.}
%\tablenotetext{c}{The ``-'' in the $N_{H}$ column indicates that the value is found to be very low and not constrained.}
\end{deluxetable}

\section{Discussion}

The X-ray emission from pulsars is typically attributed to a magnetospheric and/or a thermal origin.
Because of the small separation between the two neutron stars, a strong interaction is expected between the wind of Pulsar A and the magnetosphere of Pulsar B, inducing possibly more complex mechanisms of the X-ray emission. In the following, we discuss the origin of the soft X-ray emission (up to 4 keV) and hard X-ray emission, including the detection of the spectral feature at $6-7$ keV.

\subsection{Soft X-ray emission}

The spectral analysis of the observations performed in 2011, as well as combining 2006+2011 data, provides results comparable with those related to 2006 data alone. Different two-component (black body plus power law or double black body) or three-component spectral models (double black-body plus power law) can fit the whole dataset. However, scenarios involving only thermal emission components are disfavoured, since the flux is most likely dominated by non-thermal Pulsar A's magnetosphere emission. A double black-body model implies that the X-ray flux of Pulsar B would be about half of Pulsar A's flux (looking at black-body normalizations) and this is inconsistent with X-ray fluxes derived from timing analysis by \citet{Iacolina2016} for which X-ray emission from Pulsar A is over three times higher with respect to Pulsar B. Furthermore, pulsar phase-resolved analysis demonstrated that a non-thermal component clearly correlate with Pulsar  A's light curve peaks \citep{Pellizzoni2008}. 

No significant spectral variability in soft X-rays is evident over 5 years.
In particular, despite a significant relativistic spin precession occurred in the considered time span ($\sim 25^{\circ}$), no significant long-term variation of the soft Pulsar A's spectral component is observed.
This further supports the hypothesis of  a small misalignment between the spin axis and the orbital momentum axis as suggested  by timing analysis of the same data \citep{Iacolina2016}.

The non-thermal X-ray emission of PSR J0737-3039 is likely attributed to relativistic charged particles accelerated in the magnetosphere of Pulsar A, via synchrotron emission in the outer gap model \citep{Cheng1999} and/or via inverse Compton scattering of the thermal X-ray photons in the polar cap model \citep[and references therein]{Zhang2000}.

No signature of a bow shock component (power law with $\Gamma\,\sim\,2$ as suggested by canonical shock models; see \citealt{Sturner1997} and references therein) between the wind from Pulsar A and Pulsar B's magnetosphere is present in our soft spectral data, nor evidence of spectral variability as a function of the orbital phase. A weak orbital flux variability ($\sim$7\%) was detected in the timing analysis \citep{Iacolina2016} and it was possibly attributed to bow-shock emission. In the spectral analysis, the corresponding bow-shock flux could be concealed by the strong Pulsar A's luminosity.

% The soft photon index of the power law ($\Gamma\,\sim\,3$) indicates that there is no evidence of a bow shock between the wind from Pulsar A and Pulsar B's magnetosphere. This was already noticed by \citet{Pellizzoni2008}.
% and is corroborated by the second XMM-Newton large program.

As for the thermal emission, it is expected to originate from hot spots around the magnetic poles (polar caps) of Pulsar A and/or Pulsar B, heated up to X-ray temperatures by relativistic particles streaming down onto the surface from the magnetosphere of the pulsar(s). 
Indeed, considering the characteristic ages of Pulsar A ($ \tau_{A} = P/2 \dot{P} \sim 210$ Myr) and Pulsar B ($\tau_{B} \sim 50$ Myr),
% estimated by the spin-down rates observed today 
and the cooling evolution of such objects (Tsuruta 1998), 
these pulsars are too old to have an intrinsic thermal radiation from their entire surface or from the atmosphere \citep{Zavlin2009}. 

%Indeed, considering the characteristic ages of Pulsar A ($ \tau_{A} = P/2 \dot{P} \sim 210$ Myr) and Pulsar B ($\tau_{B} \sim 50$ Myr), estimated by the spin-down rates observed today and the cooling evolution of such objects (Tsuruta 1998), Pulsars A and B are too old to have a thermal radiation from their entire surface or from the atmosphere \citep{Zavlin2009}. 

The radio pulsar models \citep{Cheng1980,Arons1981,Beskin1993} predict polar cap radii $R_{pc} =\lbrack 2 \pi R^{3} / cP\rbrack^{1/2} \simeq 0.5\lbrack P/0.1$s\rbrack$^{-1/2}$ km for a neutron star of radius $R = 10$ km. Hence, theoretically, the polar cap radii associated with Pulsar A and Pulsar B are $\sim\,1$ km and $\sim\,100$ m, respectively. These values are in agreement with black-body emission radii derived from our spectral results.

Either assuming the presence of one or two black-body components in our spectral scenario,
the expected luminosity of each component is $> 2\times$10$^{30}$ erg/s, a value higher
than Pulsar B's spin-down luminosity. This means that the black-body component luminosity would be $> 125\%$ $\dot{E}_{rot_{B}}$. Thus, we can confirm that Pulsar B's emission can only be powered by an external source (i.e. Pulsar A's spin-down energy).

%%%%%%%%%%%%%%
%Up to five significant spectral components could be in principle expected from the Double Pulsar in X-rays accordingly to theoretical models: surface thermal emission and non-thermal magnetospheric emission from both neutron stars, and orbital-phase dependent bow-shock emission due to the interaction between Pulsar A's wind and Pulsar B's magnetosphere.  
%This latter non-thermal component, if present, was proved to be much weaker than expected: no clear evidence of shock emission was reported in the spectral analysis of 2006 data \citep{Pellizzoni2008} and the 2011 data. The spectrum appeared clearly dominated by pulsed non-thermal emission from Pulsar A and the off-pulse required two-components thermal emission models. Furthermore, only very weak orbital flux modulation ($7\%$) was detected as a possible signature of a bow-shock in the overall timing analysis of both large programs \citep{Iacolina2016}. Thus, a two (pl+bb) or three-component model (pl+bb+bb) could satisfactorily account for most of the X-ray flux from the Double Pulsar. 

%No significant spectral variations in soft X-rays can be claimed in the 5 years between 2006 and 2011, substantially confirming the previously reported results based on the analysis of the 2006 data alone. Different two-component (black body plus power law or double black body) or three-component spectral models (double black-body plus power law) can in principle fit the whole dataset, although the latter models seem better suited according to phase-resolved spectral analysis15. 

\subsection{Hard X-ray emission}

For any applicable spectral scenario in the $0.15-3$ keV range, the data above 4 keV are not correctly described by the two or three-component models (see Figure 2). 
The complementarity of the independent spatial and spectral approaches allows us to claim for the first time hard X-ray emission from the Double Pulsar. 

This hard emission is well fitted by a Gaussian line at $\sim 6-7$ keV in the 2006 data. No other simple model, such as a power law or a black body, is able to fit the high-energy excess. Emission at these energies indicates the likely presence of iron line emission. 
%The combined probability of a fake detection of hard X-ray photons through both spectral analysis and spatial analysis is negligible ($\sim 3\times 10^{-6}$).
The firm detection of hard X-ray emission in the $4-8$ keV range coupled 
with no source detection $>8$ keV through spatial analysis, strengthen the
reliability of our spectral feature detection.
The presence of such a feature has major importance since it represents the first evidence of an X-ray spectral line around a supposedly “non-accreting system”. 

%Above 8 keV, the background clearly dominates as confirmed by spatial analysis. 
As for the second long-observation performed in 2011, we cannot claim evidence of a high-energy spectral line, because of a poorer signal-to-noise ratio 
%the lower sensitivity  
in hard X-rays mostly due to the use of the thin EPIC-pn filter.
% the high-energy component appears to be different from that detected in 2006, with weaker evidence of a high-energy spectral line. However, it is worth noting that the quality of the data is not as good as during the observation performed in 2006, with a stronger background emission partly due to the use of the EPIC “thin” filter. 
%Careful simulations proved that an emission line with the same energy and flux as in the previous observation would be at the limit of detectability. 
%\textbf{We simulated fake data including a continuum (power-law plus black-body) model and a line feature with the same parameters derived from 2006 data, assuming the background model obtained from 2011 data. An emission line with the same energy and flux as in the 2006 observation would appear at the limit of detectability in 2011 data.}
In any case, the detection of high-energy photons from the Double Pulsar is also confirmed in these data through spatial analysis, though spectral parameters cannot be precisely constrained in the 2011 data alone. As described in Section 3, line variability is compatible with our data, though not required.

Associated with emission at $6.4-6.97$ keV, the Fe K$\alpha$ line is widely considered as a particular property of accretion-powered sources in which accreting matter forms a disk. It has been detected in the X-ray spectra of many supermassive black holes and stellar-mass-black-hole and neutron-star X-ray binaries \citep{Cackett2010,Miller2007}, and in magnetic cataclysmic variables \citep{Ezuka1999}. The emission line is thought to originate in the innermost parts of the accretion disk, where strong relativistic effects broaden and distort its shape. The production of such a line is fairly simple requiring a source of thermal or non-thermal hard X-rays to illuminate the disk. In this context our detection of the Fe K$\alpha$ line in the Double Pulsar may indicate that at least one of the system components is surrounded by a gaseous disk. However, what could be the origin of such a disk?

The Double Pulsar, as the other few confirmed double neutron star binaries, is the descendant of a high-mass binary system that has survived two supernova explosions \citep{Stairs2004}. The two stars under-filled their Roche lobes and lose matter at a relatively low rate through winds of relativistic particles. The conventional scenario of accretion disk formation based on intensive mass-exchange between the system components in this case is not applicable. However, a relic disk of matter captured from ejecta of the second supernova explosion could be present. The disk could be formed in two ways. The first is a so-called fall-back accretion scenario in which the inner parts of the ejecta return towards the newly-formed neutron star forming a disk around its magnetosphere \citep{Colgate1971,Zeldovich1972,Michel1988,Chevalier1989}. The mass of the remnant disk in this case is about $10^{-5}$ M$_{\sun}$. The second scenario suggests the disk to form around the magnetosphere of the old neutron star born during the first supernova explosion as it moves through the supernova ejecta of its exploded companion. As recently pointed out \citep{Bisnovatyi-Kogan2015}, the mass of the remnant disk within this scenario can reach $10^{-7}$ M$_{\sun}$. Thus, current views of the process of supernova explosions do not exclude that one or even both neutron stars could be surrounded by relic disks.

 Observations give no direct evidence that the neutron stars in the Double Pulsar are accreting material from a disk (no radio pulsar timing noise reported by \citealt{Kramer2006} and \citealt{Kramer2008}). It is plausible that the relic disk is currently in a dead state \citep{Sunyaev1977} in which mass-transfer towards the star is suppressed by the centrifugal barrier at the boundary of the stellar magnetosphere. The interaction between the disk and the magnetic field of the radio pulsar in this case results in generation of electrical currents, which connect the inner parts of the disk with polar cap regions at the surface of the neutron star \citep{Michel1981,Michel1983,Michel1985}. As the currents dissipate in small areas of the polar caps the temperature increases dramatically, enhancing the hot spots visible in X-rays. In contrast, the temperature of the disk increases little since current dissipation occurs over a much larger area. The disk in this case could contribute to the pulsar emission at lower frequencies by a significant fraction of the pulsar spin-down power.

Our spectral analysis of the XMM-Newton data demonstrates that the thermal X-ray emission is indeed incompatible with a disk origin, since the disk radius would be too small (even assuming a disk black-body model \textit{diskbb}). The surface area associated with the black bodies at 0.1 and 0.3 keV is of the order of 600 and 100 m$^{2}$, respectively, indicating most likely thermal emission from the polar cap(s). The putative disk is most likely colder and thus invisible in X-rays. It manifests itself only through the iron line that requires external heating. This could be by illumination by a hard continuum X-rays from Pulsar A, although this might be too weak or anisotropic to be observed. Alternatively, unlike accretion-powered pulsars, the disk could be illuminated by high-energy particles from Pulsar A's wind. In this case, about $0.1\%$ of spin-down energy of Pulsar A could power the K$\alpha$ line. Interestingly enough, this is the same amount of energy powering Pulsar B's X-ray emission. 

Iron line emission could also be in principle associated with an accumulation of matter trapped in the shock layer foreseen between Pulsar A's wind and the magnetosphere of Pulsar B. A hypothetical spectral line from the surface or atmosphere of the neutron star is not favored, since the gravitational redshift would induce a line at $< 6$ keV accordingly to the lower limit on the neutron stars mass-radius ratio \citep{Zhao2012}.

An idea about a relic disk to surrounding a radio pulsar has been invoked by a discovery of two planet-mass companions around the millisecond pulsar PSR 1257+12 \citep{Wolszczan1992}. Further studies eventually confirmed that the components are likely to originate from a durable remnant disk, which in a previous epoch could be in an accretion or dead state. The presence of remnant disks around isolated neutron stars has later been suspected in studies of Anomalous X-ray Pulsars (AXPs) and Soft Gamma-ray Repeaters (SGRs) \citep{Michel1985b,Chatterjee2000}. Iron line emission at 6.4 keV was detected during a gamma-ray burst in SGR 1900+14 \citep{Strohmayer2000}. The presence of a relic disk was also been invoked to explain the IR/optical radiation discovered from two AXPs \citep{Kaplan2009,Wang2006}. Only upper limits to the optical flux were derived for the Double Pulsar \citep{Ferraro2012}, which was instead detected in the FUV \citep{Durant2014}. Although the thermal or non-thermal nature of this emission is still unclear, it is not related to magnetospheric or surface X-ray pulsar emission since it does not match with extrapolated X-ray spectra. It is in principle compatible with a relic disk of luminosity L$_{FUV}=1.5 \times 10^{28}$ erg/s. 

The possible weakening or even disappearance of the spectral feature at $6-7$ keV in the data obtained in 2011 could be related to actual variability of the Fe line because of changes in the geometry or precession of the disk, which is not illuminated in the same way over time. The parameters of interaction between the wind of Pulsar A and the magnetosphere of Pulsar B are expected to be different between 2006 and 2011, as manifested by radio disappearance of Pulsar B in 2008. As the orbit is seen edge-on, changes on the disk could in principle also affect radio pulsar timing (e.g. observed dispersion measure, pulse shapes and eclipses) when it crosses the line-of-sight to the pulsars. A multi-messenger approach is then required to probe the putative relic disk and its time-dependent geometry.

\section{Conclusions}

Two Large Programs of observations of the relativistic Double Pulsar were performed by XMM-Newton in 2006 and 2011. The total observing time of about 600 ks offers the opportunity to investigate the low luminosity source with a high number of $\sim\,15,000$ source photon events.
%from the source.
%The spectral analysis of these long X-ray observations are consistent with each others and allow us to test models with an unprecedent statistics. At present, the favored model consists of a power law with a soft photon index plus a black-body component. The X-ray flux is dominated by the non-thermal emission, which is ascribed to the magnetospheric emission of Pulsar A, whereas the thermal component is consistent with the hot polar cap region of Pulsar B. This is perfectly in agreement with the timing analysis, where X-ray pulsations were detected from Pulsar B \citep{Iacolina2016,Pellizzoni2008}. A two black-body model is also a possibility, the hotter black body coming from Pulsar B, and the colder from Pulsar A. The emission radii are compatible with the polar cap radii of both pulsars estimated at $\sim\,1\,$km for Pulsar A and $\sim\,100\,$m for Pulsar B. Because of the vicinity of the two neutron stars, a more complex X-ray emission is expected from the theory, such as a bow shock between the wind of Pulsar A and the magnetosphere of Pulsar B. However, we do not see any evidence of a shock in the present data.

No significant spectral variations in soft X-rays can be claimed in the 5 years between 2006 and 2011, substantially confirming the previously reported results based on the analysis of the 2006 data alone. Two-component (black body plus power law)
%or double black body) 
or three-component spectral models (double black-body plus power law) can in principle fit the whole dataset, although the latter models seem better suited according to phase-resolved spectral analysis \citep{Pellizzoni2008}.
However, for any applicable spectral scenario in the $0.15-3$ keV range, the data above 4 keV are not correctly described by the models.

We investigated for the first time the high-energy part of the spectrum of the Double Pulsar. An intriguing emission feature, possibly variable, is detected at about $6-7$ keV.
%and is well fitted with a Gaussian line.
Spatial analysis confirms the emission of hard energy photons between 4 and 8 keV ascribed to the source. This feature is most likely attributed to iron line emission, testifying the presence of a relic disk that survived the supernova explosions that terminated the lives of the Double Pulsar's stellar progenitors. 
%A variability of this hard component is observed between 2006 and 2011, possibly ascribed to changes in the geometry of the disk between the precessing pulsars. 
%These results are also supported by the timing analysis of the same data set the Double Pulsar, with an orbital flux variability highlighted between 2006 and 2011.

\acknowledgments

We acknowledge the referee for his/her constructive suggestions that have helped to improve the content of this paper.
This work is based on observations obtained with \xmm, an ESA science mission with instruments and contributions directly funded by ESA Member States and the USA (NASA).
The authors thank the members of the \xmm\ User Support Group.
E. Egron and M. Marongiu acknowledge financial support from the Autonomous Region of Sardinia through a research grant under the program CRP-25399 PO Sardegna FSE 2007-2013, L.R. 7/2007, Promoting scientific research and innovation technology in Sardinia.
N.R. Ikhsanov acknowledges support of the Russian Scientific Foundation under the grant No. 14-50-00043.

\end{document}